# Vibration-assisted electron tunneling in $C_{140}$ single-molecule transistors

A. N. Pasupathy[1], J. Park[1], C. Chang[1], A. V. Soldatov[2], S. Lebedkin[3], R. C. Bialczak[1], J. E. Grose[1], L. A. K. Donev[1], J. P. Sethna[1], D. C. Ralph[1], and P. L. McEuen[1]

[1]Laboratory of Atomic and Solid State Physics, Cornell University, Ithaca NY 14853

[2]Physics Department, Harvard University, Cambridge MA 02138 and Department of Physics, Luleå University of Technology, SE-971 87 Luleå, Sweden

[3]Forschungszentrum Karlsruhe, Institut für Nanotechnologie, D-76021 Karlsruhe, Germany

We measure electron tunneling in single-molecule transistors made from $C_{140}$, a molecule with a mass-spring-mass geometry chosen as a model system to study electron-vibration coupling. We observe vibration-assisted tunneling at an energy corresponding to the stretching mode of $C_{140}$. Molecular modeling provides explanations for why this mode couples more strongly to electron tunneling than the other internal modes of the molecule. We make comparisons between the observed tunneling rates and those expected from the Franck-Condon model.

PACS: 73.23.Hk, 73.63.Kv, 63.22.+m, 62.25.+g

When electrons travel through molecules, vibrational modes of the molecules can affect current flow. Molecular-vibration-assisted tunneling was first measured in the 1960's using devices whose tunnel barriers contained many molecules [1]. Recently, effects of vibrations in *single* molecules have been measured using a scanning-tunneling microscope [2], single-molecule transistors [3,4] and mechanical break junctions [5]. Theoretical considerations suggest that different regimes may exist depending on whether tunneling electrons occupy resonant states on the molecule, and also on the relative magnitudes of the rate of electron flow, the vibrational frequency, and the damping rate of vibrational energy [6-14].

A quantitative analysis of electron-vibration interactions has been difficult to achieve in previous molecular-transistor experiments. In [4], neither the precise nature of the vibrational modes nor their energies was determined independently of transport measurements. In [3], the "bouncing-ball" mode of a single $C_{60}$ molecule against a gold surface was observed, a mode not intrinsic to the molecule itself. In this letter we study single-molecule transistors made using a molecule, $C_{140}$, with low-energy internal vibrational modes that are well understood. We observe clear signatures of one of these modes and discuss theoretically why it has the strongest coupling to tunneling electrons.

The $C_{140}$ molecule consists of two $C_{70}$ balls joined by two covalent C-C bonds (Fig. 1(a)). The $C_{140}$ we use was synthesized by pressure treatment of polycrystalline $C_{70}$ at 1 GPa and 200 C, purified by chromatography and characterized by $C^{13}$ NMR, Raman and infrared spectroscopy [15]. The vibrational modes of $C_{140}$ have been measured by Raman spectroscopy and modeled numerically [15]. The six lowest-energy modes are intercage vibrations in which each $C_{70}$ ball moves approximately as a rigid unit. The simple stretching mode is observed in Raman spectroscopy, with an energy of $11 \pm 0.5$ meV. The other intercage modes involve bending or twisting of the molecule, and they are predicted to be at 2.5, 2.7, 4, 15, and 17 meV. The lowest intracage excitation of $C_{70}$ is ~29 meV [16].

An atomic force microscope image of our transistor structure is shown in Fig. 1(b) and its geometry is illustrated in the inset of Fig. 1(c). A $C_{140}$ molecule bridges source and drain electrodes that are about 1 nm apart. The molecule is also capacitively coupled to a gate electrode. To make these devices, we evaporate an Al pad 20 nm thick and 2 µm wide to serve as the gate, and then oxidize in air to form the gate insulator [17,18]. On top of the gate we pattern by liftoff a gold wire 200-600 nm long, 20 nm high and 50-100 nm wide with 2-3 nm of Cr as an adhesion layer, or a platinum wire having a similar geometry without the adhesion layer. We then deposit approximately 10 µl of a 100 µM solution of $C_{140}$ molecules in *o*-dichlorobenzene onto the device area, and we allow the solvent to evaporate or we blow dry after approximately 10 minutes. After the molecules are deposited, we cool the devices to cryogenic temperatures and use electromigration to create a nm-sized gap in the wire within which a molecule is sometimes trapped [3,4,18,19]. The success rate for incorporating a molecule is approximately 10% for both the gold and platinum wires. The orientation of the molecule in the device is not known. All the measurements were performed either at 1.5 K or below 100 mK.

In Fig. 1(c), we show several current versus bias voltage (*I-V*) curves measured from a $C_{140}$ device at different gate voltages ($V_g$). The device exhibits Coulomb-blockade behavior; electron flow is suppressed at low *V* because electrons must overcome a charging energy to tunnel on or off the $C_{140}$ molecule. Plots of *dI/dV* as a function of *V* and $V_g$ are shown in Fig. 2(a) for four of the fourteen $C_{140}$ devices we have examined. The dark areas on the left and right of each plot are regions of Coulomb blockade. Tunneling can occur close to *V*=0 only near one value of gate voltage, $V_c$, which varies from device to device because of variations in the local electrostatic environment. We can be confident that the current flows through only a single molecule as long as we perform our measurements sufficiently closely to $V_g=V_c$.

The subject of this paper concerns the additional *dI/dV* lines that are observed in Fig. 2(a) at values of |*V*| larger than the boundary of the Coulomb-blockade regions. These lines correspond to thresholds involving excited quantum states of the molecule. The lines which meet the dark blockade area at $V_g < V_c$ ($V_g > V_c$) correspond to excited quantum levels of the $V_g < V_c$ ($V_g > V_c$) charge state. The energy of each level can be read off from the bias voltage where the *dI/dV* line intercepts a boundary of the blockade region (white arrows) [3].

In Fig. 2(b), we plot a histogram of all of the excited-state energies that we resolve below 20 meV in fourteen $C_{140}$ devices; excitations in each charge state are recorded separately. An excitation at 11 ± 1 meV is seen in eleven of the fourteen devices. In seven devices, the 11 meV line is present for both of the accessible charge states, while in four others it is seen for only one. In one sample (device I), well-resolved excited levels are also observed near 22 meV for both charge states, twice the 11 meV energy. As a control experiment, we measured eight devices made with $C_{70}$ molecules. No prominent peak near 11 meV is present in the histogram of $C_{70}$ levels (Fig. 2(c)).

The presence of the 11 meV excitation in $C_{140}$, but not $C_{70}$, indicates that it is an excitation of the entire molecule, and not the $C_{70}$ sub-units. The presence of the same excitation for different charge states of the same molecule, and the observation of an excitation at $2 \times 11$ meV in one device, strongly suggest that the 11 meV excitation is vibrational in nature. A purely electronic excitation should not be the same in both charge states nor should it appear as multiples of a fundamental excitation. Based on its energy, we identify this excitation with the intercage stretch mode of $C_{140}$ [15]. This is our principal result.

As shown in Figs. 2(b) and 2(c), additional excitations are present below 5 meV in both the $C_{140}$ and $C_{70}$ devices. For $C_{70}$, these are likely associated with the bouncing-ball mode of the molecule, as demonstrated previously for $C_{60}$ [3]. The sub-5-meV excitations in $C_{140}$ devices might arise either from similar bouncing modes or the intercage modes of $C_{140}$ at 2.5, 2.7 and 4 meV. However, calculations (below) suggest that the tunneling electrons couple strongly only to the intercage stretch mode, not the other internal modes. We do not observe peaks in the $C_{140}$ histogram near the bending/twisting intercage modes at 15 and 17 meV.

We will analyze our data within the framework of the Franck-Condon model [20]. $C_{140}$ has a large number of vibrational states we denote by $\mathbf{a}_j$, where $\mathbf{a}$ labels the mode of frequency $w_\mathbf{a}$ and $j$ is the number of vibrational quanta excited in the mode. For each vibrational mode, the tunneling electron drives a transition from the ground vibrational state with $A$ electrons to a vibrational state $\mathbf{a}_j$ with $B$ electrons, where $B - A = +1$ (-1) for tunneling on (off) the molecule. The tunneling rate is determined by the overlap of the starting configurational wavefunction, $\Psi_g^A$, with the one after tunneling, $\Psi_{\mathbf{a}_j}^B$:

$$\Gamma_{\mathbf{a}_j}^{A \to B} = \Gamma_{electron} P_{\mathbf{a}_j}, \text{ where } P_{\mathbf{a}_j} = \left| \langle \Psi_{\mathbf{a}_j}^B | \Psi_g^A \rangle \right|^2 \text{ and } \sum_j P_{\mathbf{a}_j} = 1. \tag{1}$$

If the electronic contribution $\Gamma_{electron}$ is assumed constant for the different vibrational transitions and if the rate-limiting step for current flow is the $A \to B$ transition, the current step associated with a given vibrational excitation is:

$$\Delta I_{\mathbf{a}_j} / \Delta I_g = P_{\mathbf{a}_j} / P_g, \tag{2}$$

where $\Delta I_g$ is the ground state current. In order to predict the size of the current steps, we must therefore calculate the atomic rearrangements that occur when a charge is added to or subtracted from the molecule. We will first perform this calculation for an isolated $C_{140}$ molecule and then discuss effects of the local electrostatic environment, before making comparisons to our measurements.

For the isolated molecule, we calculate the overlaps $P_{\mathbf{a}_j}$ using the semi-empirical method PM3 under Gaussian 03 [21]. The charge state of $C_{140}$ in our devices is not known, but since the fullerenes are easily reduced and not easily oxidized [22], we have analyzed the initial charge states $n^- = 0$, $1^-$, $2^-$, and $3^-$. The PM3 calculations indicate that the probability of tunneling without exciting any of the vibrational degrees of freedom is small. This means that tunneling at low biases is suppressed. The coupling is distributed over all of the vibrational modes, but it is large for a relatively small number. Within our measurement range ($eV < 30$ meV) the calculations indicate that the coupling is dominated by a single mode, the 11 meV stretching mode ($\alpha = s$). For the $0 \to 1^-$ transition, $P_{s1}/P_g = 0.25$. Couplings to all other vibrational modes in the measurement range are found to be smaller by at least a factor 10. The results are qualitatively similar for other charge states.

The physics of the 11 meV stretching mode can be captured using a simple model of the molecule with two masses $M/2$ connected by a spring with a spring constant $k$, as illustrated in Figure 3. The vibrational frequency is $\omega_s = (4k/M)^{1/2}$ and the zero-point rms amplitude of fluctuations in the vibrational coordinate is $x_o = [2\hbar/(M\omega)]^{1/2} = 2$ pm [20]. The length of the molecule changes by $\Delta x$ when one charge is added. The Franck-Condon result for the transition probability associated with one quantum of the stretching mode, normalized by the ground-state probability, is

$$P_{s1}/P_g = (\Delta x/x_0)^2/4. \qquad (3)$$

Higher-order transitions involving $j$ quanta of a vibrational mode have rates related to the one-quantum transitions [20]:

$$P_{s_j}/P_g = (P_{s1}/P_g)^j/j! . \qquad (4)$$

In going from the neutral to $1^-$ charge state for isolated $C_{140}$, PM3 predicts that $\Delta x = -1.9$

pm. Equation (3) then gives $I_{s1}/I_g = 0.23$, in good agreement with the full calculation above. Multiple-quanta transitions should be much smaller by Eq. (4). For other charge states, the calculated strength of the transition assisted by the stretching mode is weaker, because $\mathbf{D}x$ is smaller: for the $1^- \rightarrow 2^-$ transition $\mathbf{D}x = -0.4$ pm, for $2^- \rightarrow 3^-$ $\mathbf{D}x = -0.1$ pm, and for $3^- \rightarrow 4^-$ $\mathbf{D}x = -0.3$ pm.

The electrostatic environment in the neighborhood of the $C_{140}$ molecule may also play an important role. In general, we expect that the $C_{140}$ molecule will be subject to a strong local electric field $E$ due to image charges, work-function differences, and/or localized charged impurities. For example, an image charge at a distance 0.8 nm generates $E = 2$ V/nm. We have not succeeded in making quantitative estimates of these field-enhancement effects because the Gaussian 03 implementation of PM3 does not allow for solutions in an external field. However, a local field can be expected to preferentially enhance vibration-assisted tunneling associated with the stretching mode. When an extra electron tunnels onto $C_{140}$, the presence of $E$ will produce unequal charges on the two $C_{70}$ cages, as illustrated in Figure 3. The rearrangement of charge density within the molecule will produce changes in chemical bonding forces, leading to changes in $\mathbf{D}x$. In addition, the interaction of $E$ with the charge polarization ($+\mathbf{d}, -\mathbf{d}$ in Fig. 3) will stretch the $C_{140}$ by a length $\mathbf{D}x = E\mathbf{d}/k$. To estimate the magnitude of this stretching, assume that the charge is fully polarized: $\mathbf{d} = e/2$, and $E = 2$ V/nm. Then the electrostatic stretching is $\mathbf{D}x \sim 1$ pm. Both this stretching and the chemical-bonding rearrangement may therefore produce displacements of comparable magnitude to the values calculated above for isolated $C_{140}$.

We expect that these electric-field effects will be strongly dependent on the angle between the field and the molecular axis, because $C_{140}$ is most easily polarized along its long axis. This angular dependence means that the strength of the excited-state tunneling at 11 meV could vary significantly between devices because the orientation of the molecule in the junction may vary.

Figure 4 shows the measured ratio $DI_{s1}/DI_g$ for all of the devices. These were determined by taking the ratio of the current step height at the 11 meV peak to the current just before the step. In addition to the steps, an overall increasing background was observed that gives significant uncertainties in the step heights. In five of the devices (I, II, V, VIII, and X), the ratios were the same in both charge states, as expected within the simple Franck-Condon picture if the vibrational energies are not altered significantly by the addition of an electron. In three of these devices, $DI_{s1}/DI_g < 0.6$, consistent with the PM3 estimates above. Only the $j = 1$ vibrational state was observed for these three devices, in agreement with theory. For device I, $DI_{s1}/DI_g = 3.6 \pm 2.0$ and $2.0 \pm 0.5$ for the $n^-$ and $(n+1)^-$ state, respectively, indicating stronger coupling than expected from our estimate for isolated $C_{140}$. For this sample, additional lines were observed corresponding to the emission of two vibrational quanta ($j = 2$) with amplitudes $DI_{s2}/DI_g = 7.3 \pm 4$ for the $n^-$ charge state and $2.3 \pm 1.5$ for the $(n+1)^-$ charge state. Equation (3) predicts $6.5 \pm 4$ and $2 \pm 1$ respectively, in good agreement with the measurements. For device II, strong coupling was also observed, but no $j = 2$ line was resolved, although the increasing background may have masked its presence. Overall, then, this subset of five devices is in reasonable agreement with the Franck-Condon predictions.

In the other devices showing an 11 meV feature, unusual behavior was seen that is not expected within our simple Franck-Condon picture. Large differences were observed in $DI_{s1}/DI_g$ for the two charge states; in many cases, a line was observed only in one charge state (devices III, VI, VII, and XI). In addition, anomalously large values of $DI_{s1}/DI_g$ were observed. These could either reflect strong electron-phonon coupling or an anomalous suppression of tunneling into the ground state by vibrational or other phenomena. Pronounced negative differential resistance was present in one device (VI) [8,9].

In summary, in single-molecule transistors made from $C_{140}$ we observe vibration-assisted tunneling associated with an internal stretching mode of the molecule. The strong coupling of this mode to tunneling electrons, relative to the other molecular modes, is consistent with molecular modeling. Variations in the measured strength of vibration-assisted tunneling between different devices may be associated with an enhancement of the coupling between tunneling electrons and stretching-mode excitations by local electric fields.

We thank Anika Kinkhabwala for experimental help and Karsten Flensberg for discussions. This work was supported by the NSF through the Cornell Center for Materials Research DMR-0079992, DMR-0244713, ACIR-0085969, and use of the NNUN/Cornell Nanofabrication Facility.

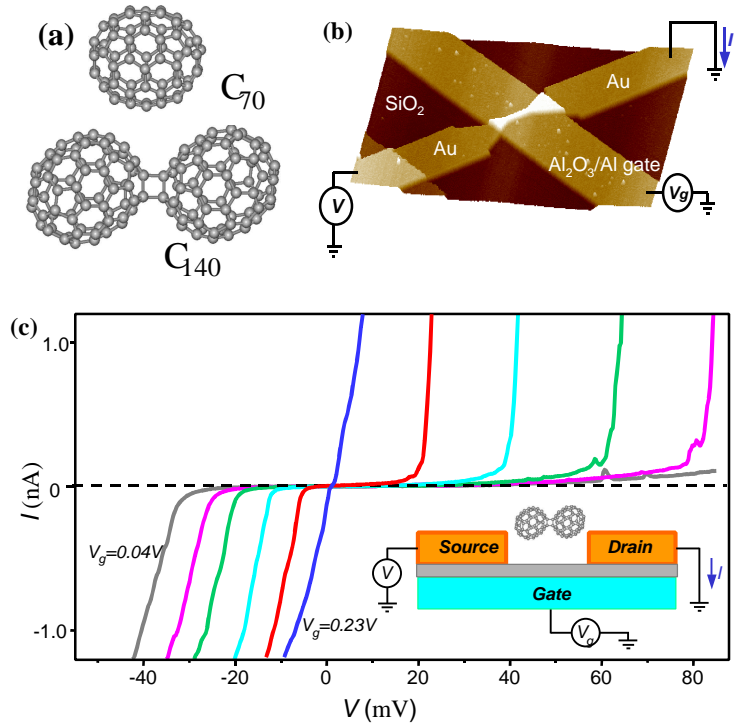

FIG 1. (a) $C_{70}$ and $C_{140}$. (b) An atomic force microscope image of a continuous Au electrode, before we perform electromigration, fabricated on top of an oxidized Al gate electrode. (c) *I-V*'s from a $C_{140}$ single electron transistor for equally-spaced values of $V_g$. Inset: schematic of the device geometry.

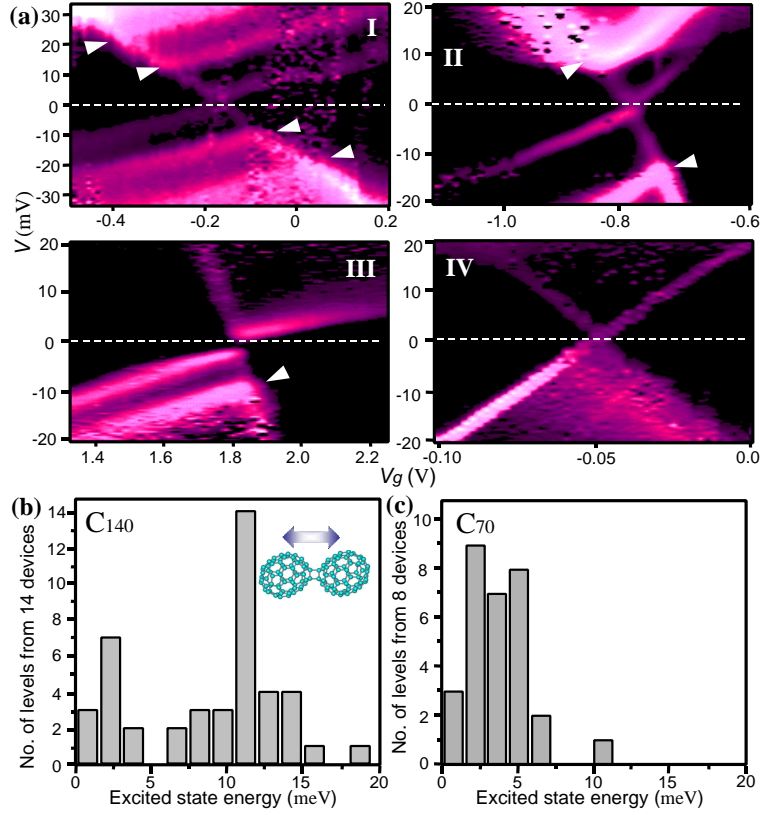

**FIG. 2.** (a) *dI/dV* vs. *V* and $V_g$ for four $C_{140}$ devices. White arrows indicate excited levels at 11 meV and 22 meV. We include examples where the 11-meV levels are observed only in one charge state (III) and in neither state (IV). *dI/dV* is represented by a color scale from black (zero) to white (maximum), with maximum values 200 nS (device I), 600 nS (II), 15 nS (III), and 100 nS (IV). Measurements were done at 1.5 K for I-III and 100 mK for IV. (b) A histogram of observed excited energies from 28 charge states in 14 $C_{140}$ devices. (c) A similar histogram for 8 $C_{70}$ devices.

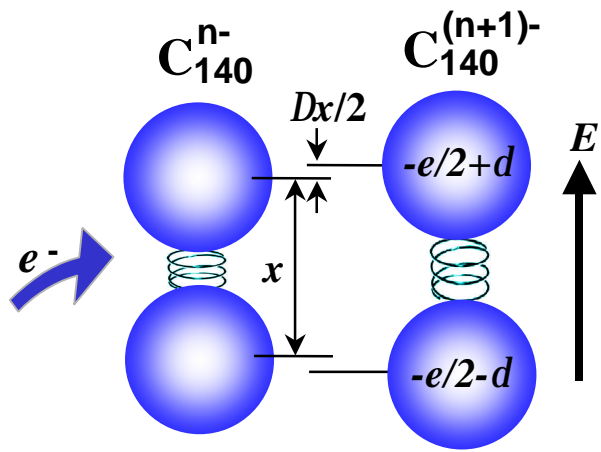

**FIG. 3.** One possible mechanism for the field-enhanced excitation of the intercage vibrational mode.

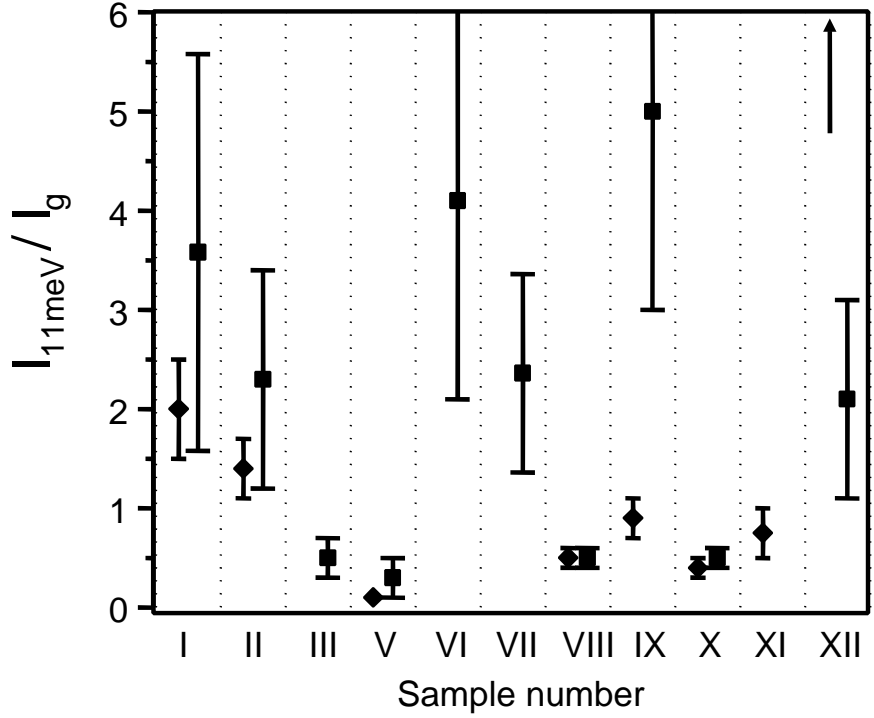

FIG. 4. Values of $I_{11\text{meV}}/I_g$, the measured current step for the excited-state signal relative to the ground-state current. Values for both charge states $n^-$ (squares) and $(n+1)^-$ (diamonds) are shown. One value is not displayed: for device XII, $I_{11\text{meV}}/I_g = 8 \pm 2.5$ for the $(n+1)^-$ charge state. Samples IV, XIII, and XIV have no visible 11-meV levels.